\documentclass{article}
\usepackage[letterpaper,top=2cm,bottom=2cm,left=3cm,right=3cm,marginparwidth=1.75cm]{geometry}
\usepackage[dvipsnames]{xcolor}

\usepackage{nicematrix} %\begin{NiceTabulat}
\usepackage{braket}
\usepackage{hhline} % \hhline{=|=|=}

\definecolor{orange}{cmyk}{0, 0.7, 0.7, 0.5}

\usepackage{amsmath}
\usepackage{graphicx}
\usepackage[colorlinks=true, allcolors=blue]{hyperref}
\usepackage{booktabs}
\usepackage{siunitx}
\DeclareSIUnit\clight{\text{\ensuremath{c}}}

\usepackage[english]{babel}

\usepackage{biblatex}
\DefineBibliographyExtras{american}{\stdpunctuation}
\title{%LCWS 2021: Proceeding \\ 
Designing a plasma lens as a matching device for the ILC positron source\footnote{Talks presented at the International Workshop on Future
  Linear Colliders (LCWS2021),\\
  \mbox{}\hspace{5mm} 15-18 March 2021. C21-03-15.1} }
\author{M. Formela, N. Hamann, K. Fl\"ottmann, G. Moortgat-Pick, S. Riemann}

\begin{document}
\maketitle

\begin{abstract}
To realise a planned high-luminosity and high-energy $e^+e^-$-collider, as the ILC, a large amount of positrons have to be produced and the accelerated particles have to be captured and matched according to the damping ring acceptances. 
%There exist several technical possibilities. 
In this contribution a new promising alternative method for capturing positrons will be presented, the application of the plasma lens as an optical matching device. It will be compared with the current matching device proposed for the ILC, namely the quarter wave transformer. An advantage of the plasma lens is the different magnetic field component, which focuses the divergent beam in a more effective manner. Therefore it will be shown in this paper that the yield requirements could be achieved more easily. The plasma lens can actually be a promising alternative for focusing beams as soon as the technical feasibility has been approved.\\
In the simulation, a tapered active plasma lens has been optimized using the approximation of a homogeneous electric current density constant in time. The optimization process led to a plasma lens design that improves on the ILC's currently proposed optical matching device, namely the quarter wave transformer, by approximately $50-100\,\%$. Furthermore the design has been shown to guarantee a stable captured positron yield within $\pm1.5\%$ for single, independent parameter deviations of about $\pm10\%$.
\end{abstract}

\section{Introduction}
\begin{figure}[!htbp]
\centering
\includegraphics[width=0.5\textwidth]{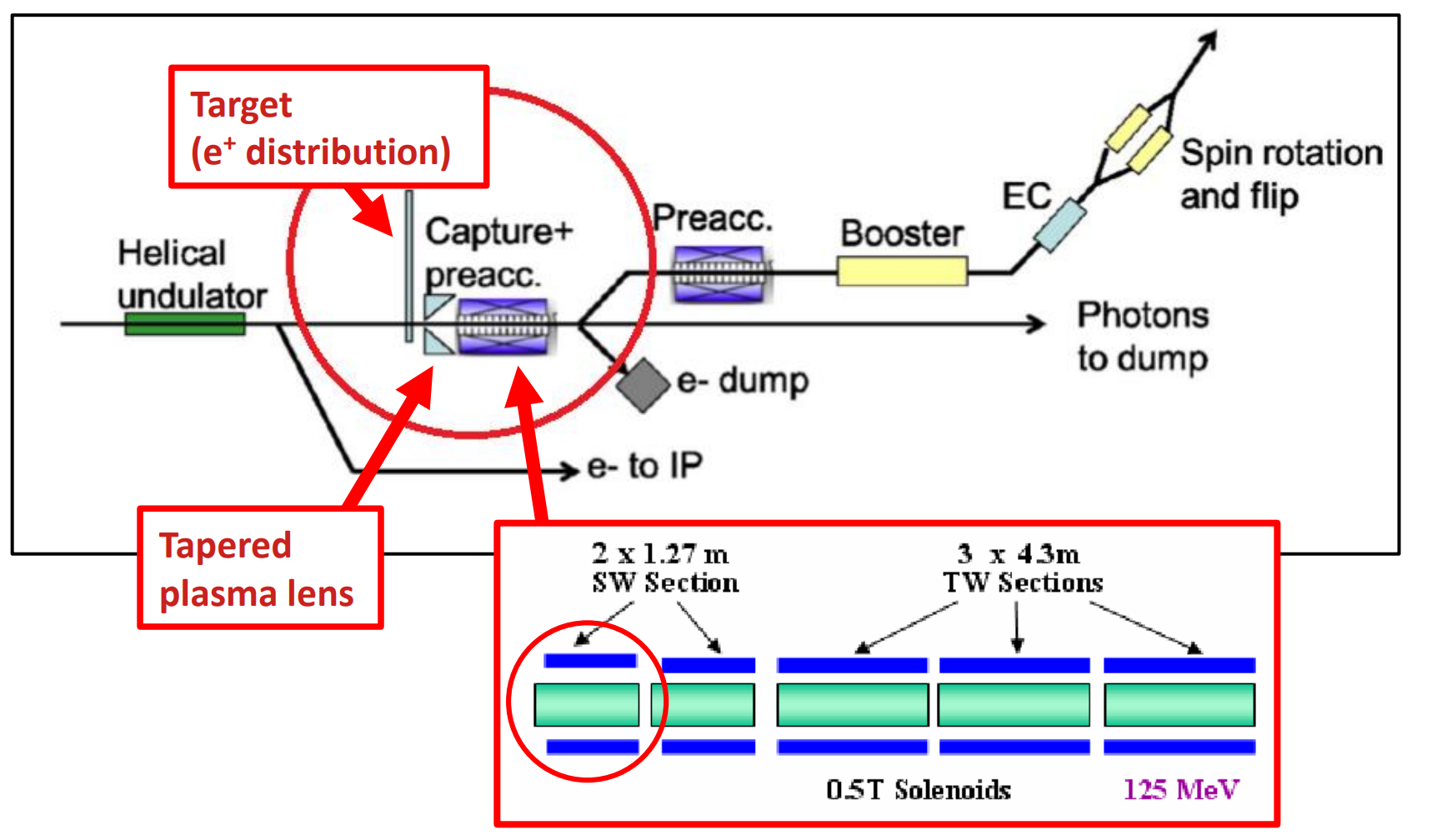}
\caption{\label{fig:Sketch} A sketch of the ILC components which are used in the simulations. Also included, the pre-accelerator and capture structure. Only the first standing wave tube is used. Source ILC Sketch: \cite{riemann2020updated}. Source pre-accelerator structure: \cite{wang2007positron}}
\end{figure}

A challenge for all future high-energy $e^+e^-$-collider designs is the required 
high positron production rate.
%If one wants to build a new particle collider one challenge is the production of particles, 
In case of the International Linear Collider (ILC), 
it is currently proposed to use a quarter wave transformer as optical matching devices (OMD) for capturing and matching of the positrons. However, for the recently fixed starting energy stage of $\sqrt{s}=250$~GeV, the proposed OMD design is not optimized and further promising alternatives are appreciated. 
%The goal of this paper is to study the tapered plasma lens as new possible alternative.  \\
The tapered plasma lens proposed in this paper is labeled as an adiabatic matching device (AMD), which means that the magnetic field used for focusing is decreasing along the z-axis. This should result in a parallel beam after passing through the device. Tapering refers to the changing radius of the plasma lens along the z-axis. Hereafter the term `plasma lens' is used interchangeably for a tapered plasma lens.\\
The simulations discussed in this paper are done via the ASTRA code \cite{floettmann2017astra}. They consist of the initial positron distribution privately commissioned by M. Fukuda, the plasma lens and the capture region or pre-accelerator structure which includes a constant magnetic field of $0.52\,\mathrm{T}$. A sketch of the structure can be seen in Figure \ref{fig:Sketch}. Important to notice is that in these simulations only the first Standing Wave Tube (SWT) of the pre-accelerator region is implemented. This is done to save calculating time, because the first standing wave tube serves well as a first approximation. The standing wave tube is in total $1.27\,\mathrm{m}$ long. Also important is that all simulations are done without considering space charge.
\newpage

\section{Initial Positron Beam}
\begin{figure}[!htpb]
    \centering
    \includegraphics[width=\textwidth]{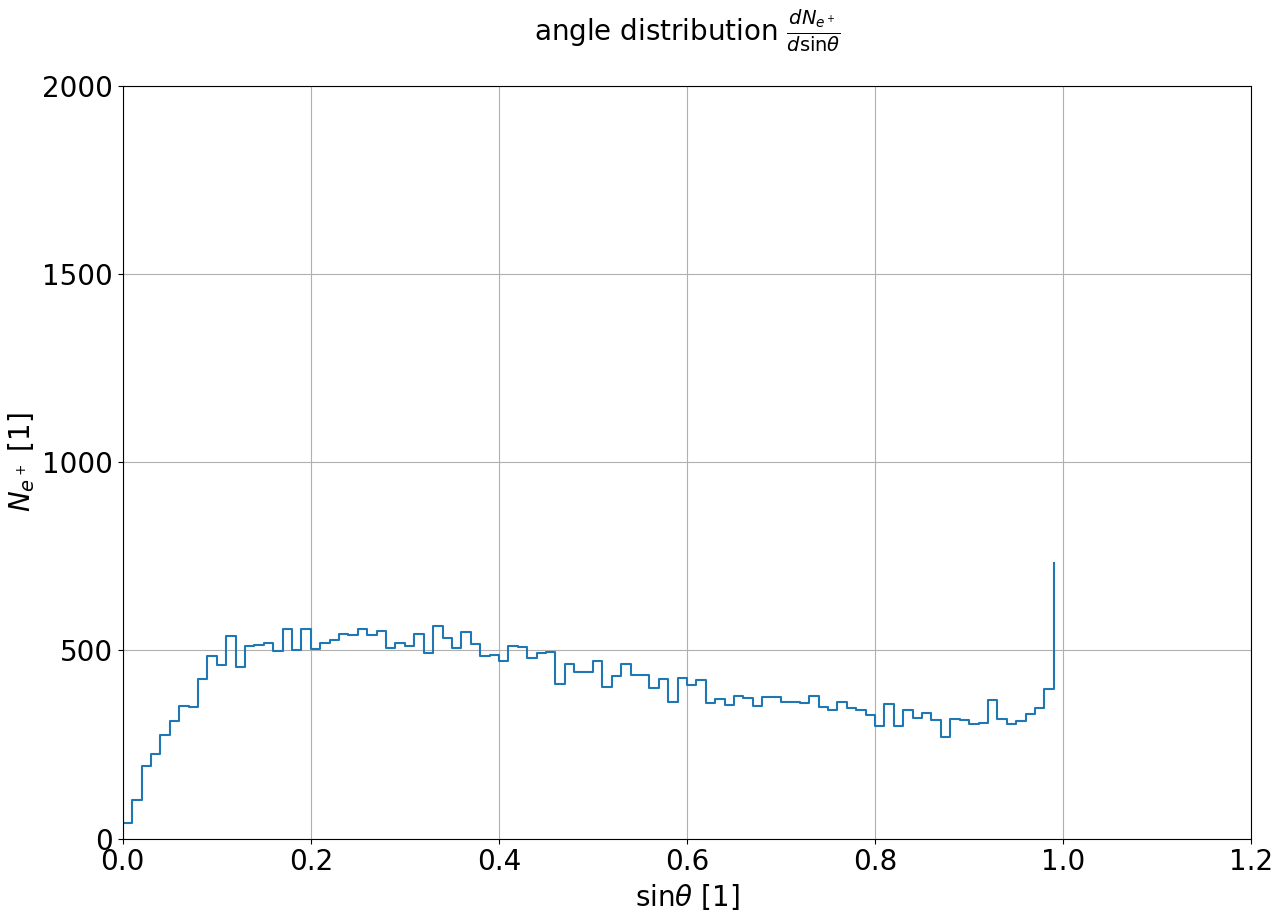}
    \caption{Opening angle $\sin\theta$ distribution. Bin size: $0.01$.}
    \label{fig:theta_distr}
\end{figure}

\begin{figure}[!htpb]
    \centering
    \includegraphics[width=\textwidth]{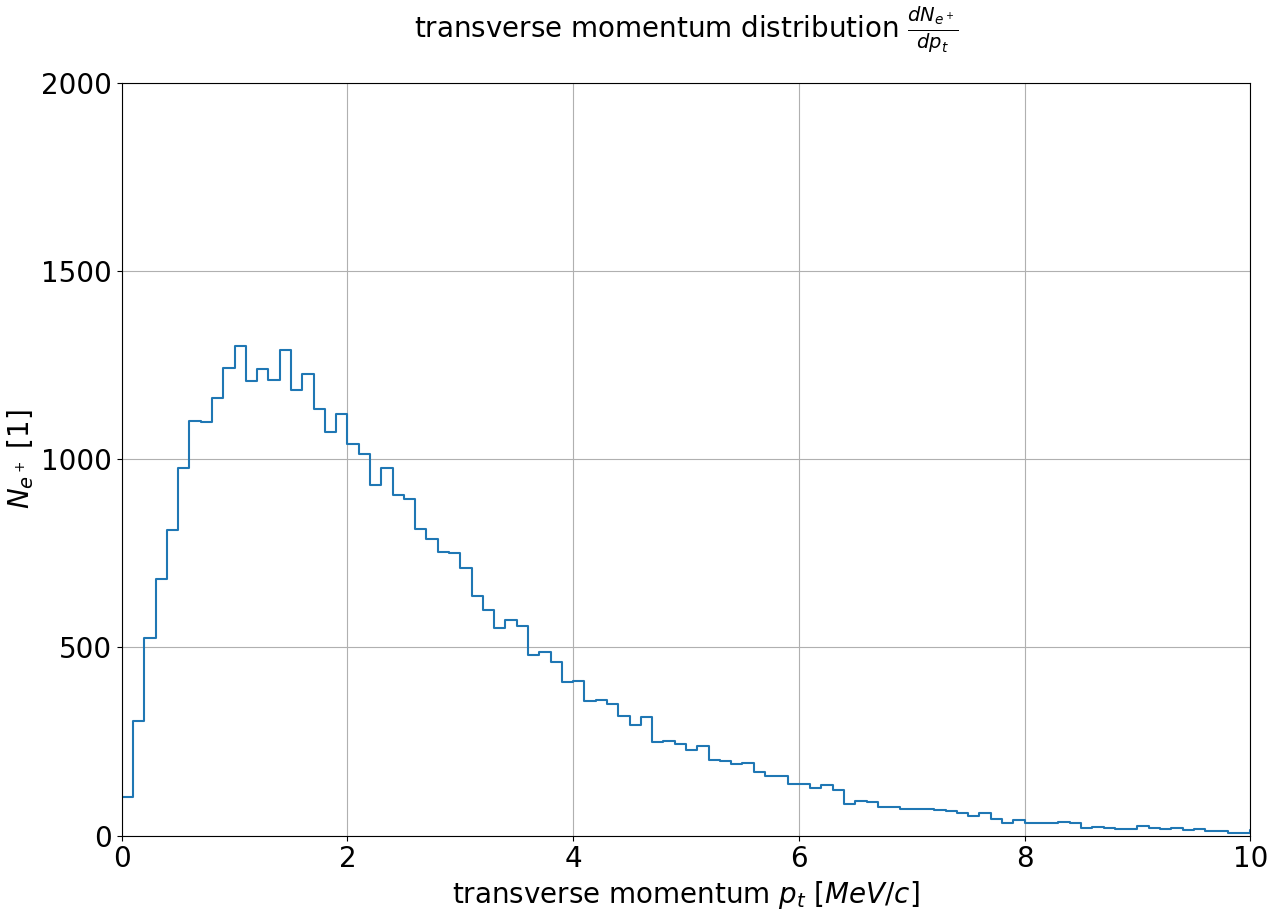}
    \caption{Transverse momentum $p_t$ distribution. Bin size: $\SI{0.1}{\MeV\per\clight}$.}
    \label{fig:pt_distr}
\end{figure}

\begin{figure}[!htpb]
    \centering
    \includegraphics[width=\textwidth]{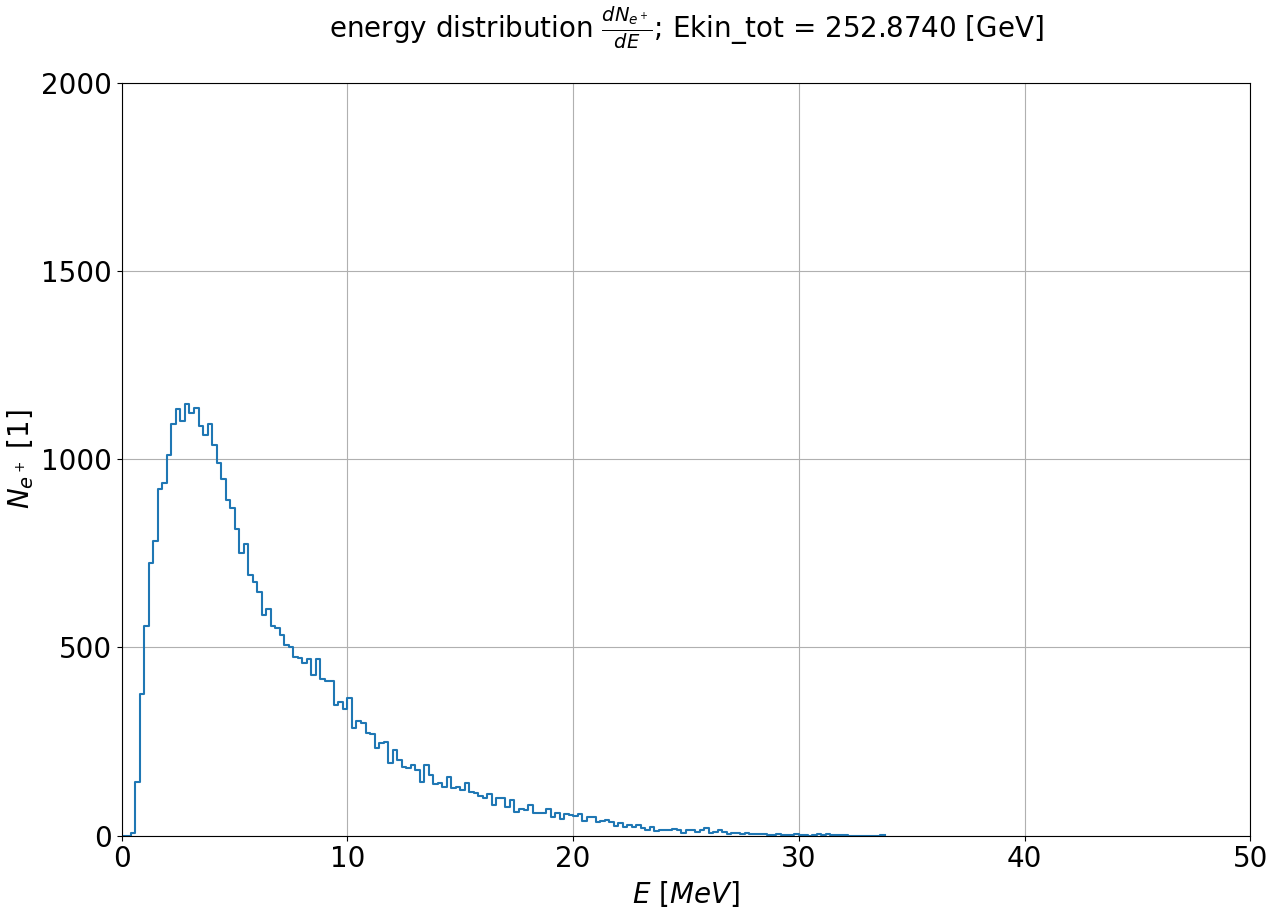}
    \caption{Kinetic energy $E_{kin}$ distribution. Bin size: $\SI{0.2}{\MeV}$.}
    \label{fig:Ekin_distr}
\end{figure}

The positron distribution stems from simulations of ILC's proposed undulator-driven scheme with a $\SI{7}{\milli\meter}$ thick Ti6Al4V target and was generously provided and calculated by M. Fukuda and K. Yokoya from KEK. In the latter study,  $10,000$ electrons were simulated at $\SI{125}{\GeV}$ generating $4,025,930$ photons in CAIN, which was used as input by the former study to calculate the positron distribution in GEANT4.\\
The tracked initial particle distribution consists of $41,564$ forward travelling positrons forming a short, divergent and small bunch. Backwards travelling particles and particles types other than positrons (e.g. electrons from pair production) are neglected. The exact bunch parameters are shown in Table \ref{tab:ini_beam_para}.
\begin{table}[ht]
    \centering
    \begin{NiceTabular}{l||l|c|c}
        ~~~~~~~~~~~~~~~~~~~~~~~~~~~ & \textbf{Parameter name} & \textbf{Symbol} & \textbf{Value} \\ \hhline{=#=|=|=}
        
        \Block{2-1}{Statistics:} & Number of positrons & $N_{e^+}$ & $41564$ \\
            & Number of electrons & $N_{e^-}$ & $0$ \\ \hline
        \Block{4-1}{Spatial Distribution:} & Average x position & $\braket x$ & $\SI{0.0}{\mm}$ \\
            & Average y position & $\braket y$ & $\SI{0.0}{\mm}$ \\
            & Beam size (rms) & $\sigma_{x,rms}$ & $\SI{1.62}{\mm}$ \\
            & Beam size (rms) & $\sigma_{y,rms}$ & $\SI{1.34}{\mm}$ \\ \hline
        \Block{2-1}{Time Distribution:} & Average emission time & $\braket t$ & $\SI{82.07}{\pico\second}$ \\ 
            & Emission time spread (std)  & $\sigma_{t,std}$  & $\SI{2.28}{\pico\second}$  \\ \hline
        \Block{2-1}{Energy distribution:} & Average kinetic Energy & $\braket{E_{kin}}$ & $\SI{6.08}{\mega\eV}$ \\ 
            & Energy spread &  & $\SI{4.82}{\mega\eV}$ \\ \hline
        \Block{4-1}{Emittance:} & Normalized x emittance (rms) & $\epsilon_{x,n,rms}$ & $\SI{6914.9}{\pi\milli\radian\mm}$ \\
            & Normalized y emittance (rms)& $\epsilon_{y,n,rms}$ & $\SI{5731.8}{\pi\milli\radian\mm}$ \\
            & Correlated x divergence &  & $\SI{59.52}{\mm}$ \\
            & Correlated y divergence &  & $\SI{71.75}{\mm}$ \\ \hline
     \end{NiceTabular}
    \caption{Initial positron distribution parameters (undulator driven at $\SI{125}{\GeV}$; $\SI{7}{\milli\meter}$ thick Ti6Al4V target).}
    \label{tab:ini_beam_para}
\end{table}\\
Further figures are shown in the Appendix: the transverse dimensions (Fig.~\ref{fig:xy_distr}), the 
angle distribution (Fig.~\ref{fig:theta_distr}), the transverse momentum distribution (Fig.~\ref{fig:pt_distr}) and the energy distribution (Fig.~\ref{fig:Ekin_distr}).

\section{Operation Cycle of an Active Plasma Lens}
An active plasma lens (PL) consists of an open cylindrical cavity (capillary) with ring electrodes around both ends and gas inlets. The capillary's symmetry axis coincides with the beam axis. \\
In the first step gas (e.g. $H_2$) is introduced into the capillary, followed by applying some $\SI{}{\kV}$ pulsed voltage to the ring electrodes and producing a strong longitudinal electric field inside the capillary. In fact the field is of such magnitude that the neutral gas atoms are ionized and form an ion-electron mixture, the so called plasma. Whereas the ions momenta are virtually unchanged as a result of their high rest mass, the electrons, however, are accelerated in longitudinal direction by the present electric field, forming a short-lived electric discharge current of some $\SI{}{\kA}$. In turn the discharge current in longitudinal direction generates an azimuthal magnetic field. Inserting a single charged particle bunch into the PL leads then to a radial force, resulting in a focused bunch exiting the PL. Depending on the discharge current's life time, which has an upper bound defined by the high voltage pulse, multiple bunches or even a whole particle pulse can be transformed. Afterwards only ions are left inside the capillary, which then disperse through the openings into the vacuum. This marks the end of the duty cycle and a new one can be initiated.

\section{Issues of Conventional Optical Matching Devices}
The optical matching device (OMD) resides right between the target and the pre-accelerator section and it is responsible for transforming the particles exiting the target from a highly divergent beam with a small effective cross-section to a wide, parallel one, whose phase-space distribution matches appropriately the succeeding accelerator sections. \\
In the past this problem has been approached by different types of sophisticated coils like the quarter wave transformer (QWT) and flux concentrator (FC). Both have fundamentally a problem with strong dephasing. The QWT and the FC also suffer from chromaticity and eddy currents in rotating targets, respectively. The plasma lens as a new alternative OMD option could have fundamentally less issues in all these three areas.

\subsection{Dephasing}
The dephasing is concerned with the longitudinal dynamics of the particle bunch within the OMD. The longitudinal size is primarily determined by the effective length of the trajectory of the bunch on which the focusing magnetic field acts on. 
Both the QWT and the FC utilize currents flowing circularly around the beam axis and therefore produce a primarily longitudinal magnetic field, which results in a helical trajectory. Unlike the plasma lens with its longitudinal discharge current producing an azimuthal magnetic field and therefore leading to a sinusoidal trajectory. In principle, a sinusoidal path is favourable 
%over the one of a helix 
due to the shorter effective path the particles traverse, giving the bunch less time to fly apart due to the 
bunch's
wide energy spread pre-acceleration.
In Figures~\ref{fig:Cut_PL} and \ref{fig:Cut_QWT} the difference in degree of the dephasing can be observed. Both plots show the longitudinal particle distribution with information on the momentum deviation. Figure~\ref{fig:Cut_PL} belongs to the plasma lens and Figure~\ref{fig:Cut_QWT} to the quarter wave transformer. The red lines indicate the longitudinal cut (see section~\ref{sec:DR acc}), which highlights the fact, that the positron yield is higher for the plasma lens due to the different dephasing behaviour. As one can see, the plasma lens provides a compact, focused particle bunch, compared to the much wider spread bunch of the quarter wave transformer. In this regard, one would also expect an 
%equally (FC ist vielleicht noch schlimmer)
unfavourable behaviour of the flux concentrator.

\subsection{Chromaticity}
The chromaticity is concerned with the energy dependence of the focusing process. In general, particles with energies deviating from the design energy will experience an off-design focusing. Or in other words, the focal length is an energy dependent quantity, limiting the effectiveness of the process.
The chromaticity of the quarter wave transformer is high, while the flux concentrator and the plasma lens offers smaller energy dependence. A more detailed explanation can be seen in section~\ref{subsec:QWT thin lens}.

\subsection{Eddy Currents}
Due to the OMD's natural proximity to the target, the effects and risks of edge fields spilling into a rotating, conductive target have to be taken into account. Indeed, a magnetic field perpendicular to the target's motion induces an eddy current in the latter, leading to the exertion of a counteractive drag force. This results in stress on the target's propulsion with potential repercussion on its lifetime or an immediate malfunction. \\
As the magnetic fields 
of the QWT and the FC
are not dissimilar to that of a solenoid, their longitudinal fields within the device reach out to form closed loops. This means that edge fields will inevitably penetrate the target and the aforementioned considerations become relevant. For the quarter wave transformer the effects are manageable, whereas this is more problematic for the flux concentrator. The plasma lens, on the other hand, avoids this issue altogether by its azimuthal magnetic field lines, which form closed loops in the device's interior and are limited to it.

\section{Trace-Space Transformation}
\begin{figure}[ht]
    \centering
    \includegraphics[width=1.0\textwidth]{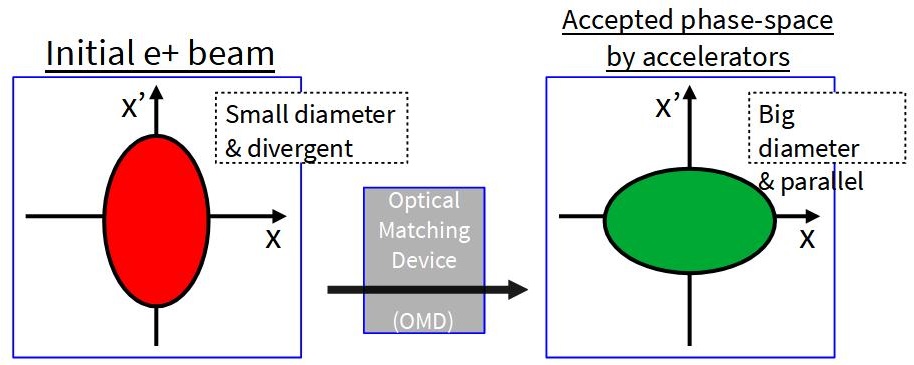}
    \caption{The OMD's general task of transforming the trace-space ellipse.}
    \label{fig:omd_principle}
\end{figure}

The task of an optical matching device is to match the transverse dimensions and the divergence of a particle distribution to the acceptance of the succeeding sections, i.e. the damping ring. This can also be expressed in terms of the phase- or trace-space. In this case, we are interested in the 4-dimensional trace-space consisting of two two-dimensional trace-planes $(x,x')$ and $(y,y')$ with the transverse coordinates $x$, $y$ on the horizontal axis and the corresponding transverse momentum ratios $x'=\frac{p_x}{p_z}$, $y'=\frac{p_y}{p_z}$ as the vertical axis. If we draw every particle of the initial bunch as points into these coordinate systems, we get a distribution resembling an up-right ellipse. The ellipse is elongated vertically, because its divergence is high, and compressed horizontally, because its transverse size is small. If we now draw the same Figure for particles, which just fulfill exactly the damping ring requirements, we receive an ellipse, that is rotated relative to the previous one by $90$ degrees. This ellipse shows the trace-space of an wide and parallel distribution. Now it is possible to phrase the task of an optical matching device in terms of the trace-space: The optical matching device is responsible for moving the particles in the trace-space in such a way, so that at the end as many particles as possible occupy the space within the target ellipse (see Figure~\ref{fig:omd_principle}). There are several ways to achieve this. In the following the principles of the quarter wave transformer and of an adiabatic matching device are discussed. 
\subsection{Quarter Wave Transformer in Thin Lens Approximation.} \label{subsec:QWT thin lens}
\begin{figure}[ht]
    \centering
    \includegraphics[width=1.0\textwidth]{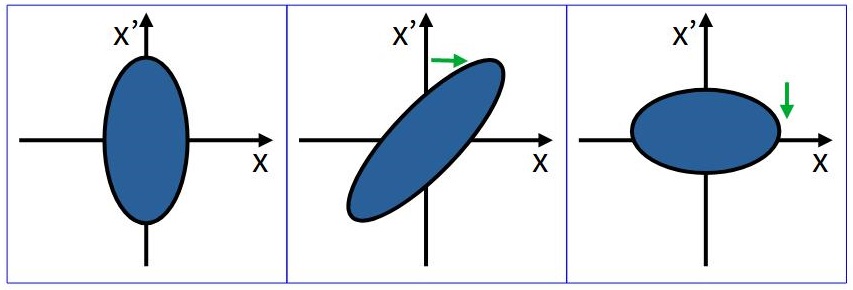}
    \caption{Transformation process of the trace-space ellipse for a quarter wave transformer. In the case of a single particle energy. Left: initial ellipse. Middle: after drift section. Right: after quarter wave transformer.}
    \label{fig:qwt_one_energy}
\end{figure}

\begin{figure}[ht]
    \centering
    \includegraphics[width=1.0\textwidth]{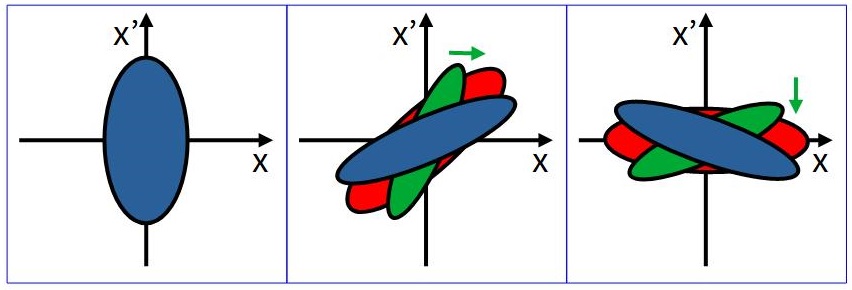}
    \caption{Transformation process of the trace-space ellipse for a quarter wave transformer. In the case of three different energy classes of particles. Left: initial ellipse. Middle: after drift section. Right: after QWT.}
    \label{fig:qwt_wide_energy}
\end{figure}

The quarter wave transformer in the thin lens approximation achieves the proper trace-space transformation by skewing the initial ellipse in two steps. The following explanation is visualized in Figure \ref{fig:qwt_one_energy}.\\
In a first approximation we assume that the initial distribution consists of particles of identical energy. Letting this simple initial beam  pass undisturbed through a drift section has consequences on the initial up-right ellipse. Namely, the ellipse is skewed in horizontal direction around the ellipse center. This transformation of the whole ellipse, can be explained by looking at the individual particles from which it is made of. Particles further away from the horizontal axis, namely with higher transverse momentum, 
wander in a straight line further away from the vertical axis. Respectively, particles closer to the horizontal axis, namely particles with smaller transverse momentum, wander in a straight line only a little bit away from the vertical axis. In fact, particles living exactly on the horizontal axis stay unmoved. Looking at the whole distribution this is equivalent to the beam widening up due to its natural divergence, while the divergence itself is unchanged due to the lack of external forces. Afterwards only one more skewing process is necessary to finalize the matching process. The necessary skew direction is vertical and is conducted by the quarter wave transformer. In analogy to the horizontal skewing process, a single particle moves more in horizontal direction in trace-space the further away it is located from the vertical axis, 
i.e.\ when its distance from the beam line is large. This is equivalent to lowering the absolute value of the divergence. 
%namely the paralleling of the bunch.
 \\
This approach is only effective as long as the initial particle distribution has a narrow energy width. This can be made plausible by going through the matching process while considering additional particles from two more energy classes, one above and one below the energy of the previous discussion. The initial trace-space ellipse looks identical in both cases. The effect only becomes apparent after the drift section. Instead of one single skewed ellipse, three distinct ellipses can be seen, one for each particle energy class (see Figure \ref{fig:qwt_wide_energy}). This is due to the fact that the difference in energy results in different degrees of skewing. The distance a particle traverses through the trace-space during a drift section is not only dependent on the transverse momentum ratio, but also on the particles energy. In fact, the correlation is linear. Therefore ellipses of lower energy particles are skewed less than those of high energy particles. The result is the overlapping of differently skewed ellipses, which will not be fixed by the 
%following 
quarter wave transformer. Instead the previous under- and over-skewing is passed on and therefore particles with energies deviating from the design energy are not matched properly. This makes the QWT a narrow energy-band device.
\subsection{Adiabatic Matching Device}
\begin{figure}[ht]
    \centering
    \includegraphics[width=1.0\textwidth]{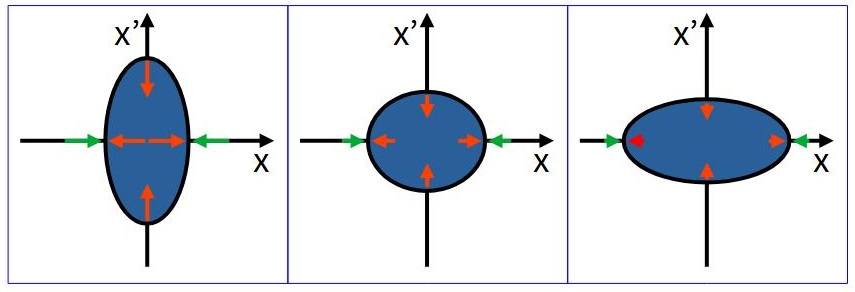}
    \caption{Transformation process of the trace-space ellipse for a adiabatic matching device. Left: initial ellipse. Middle: inside the AMD. Right: at end of AMD.}
    \label{fig:amd}
\end{figure}

Adiabatic matching devices, like the flux concentrator and the tapered plasma lens, transform the trace-space ellipse in one continues process, in an adiabatic expansion.
During the adiabatic expansion, the up-right trace-space ellipse is being squashed vertically and stretched horizontally until the target ellipse is covered optimally (see Figure \ref{fig:amd}). 
 \\
It is important to note, that this is less sensitive to the energy width of a particle bunch. This is in contrast to the quarter wave transformer and makes the adiabatic matching device a broad energy-band device. This advantage, however, comes with an drawback: the adiabatic expansion requires a slowly tapered magnetic field and therefore a long device. This in turn allows the longitudinal dynamics, the so called dephasing, to play a bigger role, especially when considering the helical particle trajectory in the flux concentrator. The particle bunch disperses more, making the longitudinal acceptance harder to fulfill, so that in the end one has to strike a balance between better transverse and worse longitudinal matching.

\section{Damping Ring Acceptances} \label{sec:DR acc}
\begin{figure}[!htbp]
\centering
\includegraphics[width=0.5\textwidth]{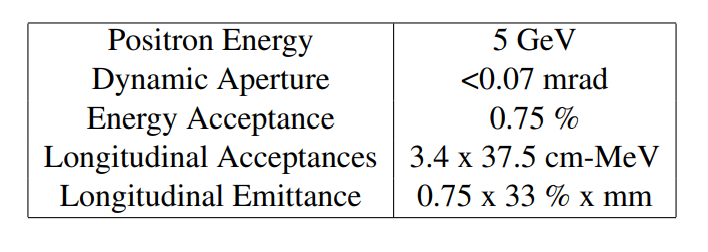}
\caption{\label{fig:DR Acc} ILC damping ring acceptance. \cite{adolphsen2013international} }
\end{figure}

For accurate simulations many conditions have to be fulfilled. One of the most important conditions are bound to the damping ring. Therefore it is necessary to work out  the limitations of the beam structure after the optical matching device imposed by the damping ring. In Figure~\ref{fig:DR Acc}, one can see the damping ring acceptances which are used to determine the final state simulations. As one can see the pre-accelerator structure is designed to accelerate the positrons to a final energy of $5\,\mathrm{GeV}$ with an energy acceptance of $0.75\,\%$ \cite{adolphsen2013international}. The final energy is not that important as the actual energy acceptance. This energy acceptance leads to a longitudinal acceptance of $\pm 7\,\mathrm{mm}$, i.e.\ to an accepted bunch length in total of $14\,\mathrm{mm}$. Therefore the particle distribution along the z-axis should be as dense as possible. Since these simulations are a first study for designing a plasma lens as an optical matching device for the ILC positron source, only the energy acceptance will be considered. This means a cut after the simulations will be done by hand. With this cut it will be ensured that only positrons which reside within the $14\,\mathrm{mm}$ longitudinal acceptance of the damping ring will be considered active particles. This cut will be named longitudinal cut from here on after. A more comprehensive study including also other possible cuts will be done in future.

\section{Optimization of Parameters}
\begin{figure}[!htbp]
\centering
\includegraphics[width=0.4\textwidth]{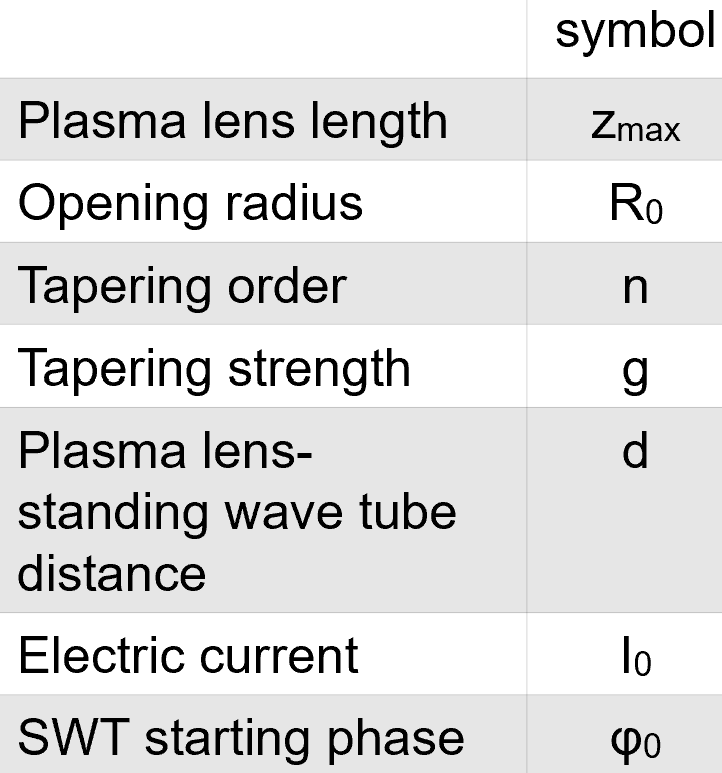}
\caption{\label{fig:Para} Parameters for the plasma lens which can be optimized. }
\end{figure}

The goal is to find a possible alternative optic matching device to the proposed QWT, as mentioned before. Therefore it is important to find a design of the plasma lens in regards of a maximal positron yield after the longitudinal cut, meaning the whole optimization process is focused on maximizing the captured 
number of positrons.
%In order to  optimize, 
A few parameters have to be varied to obtain the best result. Some of these parameters are generated by introducing an equation for calculating the changing aperture radius of the tapered plasma lens, see 
Eqn.~(\ref{Eq: Ap radius}). In the following the relevant parameters are shortly described.
First of all there is the total length of the plasma lens denoted as $\mathrm{z_{max}}$.
The opening radius of the plasma lens, called $\mathrm{R_{0}}$,  is also important for calculating the aperture radius. In Eqn.~(\ref{Eq: Ap radius}), one can also see the tapering order $\mathrm{n}$, which changes the overall shape of the plasma lens and therefore also the magnetic field. 
The tapering strength, $\mathrm{g}$,  describes how strong the magnetic field decreases along the z-axis, implying that the radius of the plasma lens has to increase along the z-axis. 
Furthermore, 
the distance between the plasma lens and the standing wave structure $\mathrm{d}$, the current strength of the plasma lens which also defines the strength of the magnetic field $\mathrm{I_{0}}$ and last but not least the starting phase of the standing wave tube $\mathrm{\Phi_{0}}$ have to be taken into account, see Figure~\ref{fig:Para} summarizing the optimized parameters.
\begin{equation}  \label{Eq: Ap radius}
    R(z) = R_{0}(1+gz)^{n}.
\end{equation}

\newpage

\section{Optimization Process}
The process of searching for the best possible parameters is time consuming, only one parameter will be changed while all other will stay constant. After finding the maximum for the changed parameter. This ideal parameter will be then set to be constant and the next parameter will be changed. 
%Because in previous designs the current strength is set to be too large in a realistic way 
%one main goal is to have a current strength as low as possible.
%For that reason at 
In order to have a rather low current strength, we fix it
%the start of the optimization the current strength is set 
to $3000\,\mathrm{A}$, which eliminates this parameter for further optimization.
\subsection{Results}
After optimizing all these parameters a final result can be calculated and the ideal parameters can be seen in Table \ref{tab:optimal_parameters-PL}.
After applying the longitudinal cut of $14\,\mathrm{mm}$ we get a positron yield of about ${\bf 41.7\,\%}$ with a total of $42917$ initial positrons. 

\begin{table}[!htbp]
   \centering
   \begin{tabular}{lcc}
       \toprule
       \textbf{Parameter name} & \textbf{Symbol} & \textbf{Optimal Value} \\
       \midrule
    Plasma Lens Length & $z_{max}$ & $\SI{6}{\cm}$ \\
    Opening Radius & $R_\mathrm{0}$ & $\SI{3.8}{\mm}$ \\
    Tapering Order & $n$ & $1$ \\
    Tapering Strength & $g$ & $\SI{136}{\m^{-1}}$ \\
    PL-SWT distance & $d$ & $\SI{1}{\cm}$ \\
    SWT Phase & $\varphi_\mathrm{0}$ & $\SI{220}{\deg}$ \\
       \bottomrule
   \end{tabular}
    \caption{Ideal parameters with the best possible positron yield at the electric current $I_\mathrm{0}=\SI{3}{\kA}$.}
   \label{tab:optimal_parameters-PL}
\end{table}

\subsection{Stability}
%Finding the ideal parameters is one step. Another
The next important 
step is to check whether these parameters are sensitive to variations or not, for instance, 
%This was done to see 
whether maladjustment of the plasma lens are compromising the yield in a negative manner. 
%For this one can do 
Therefore we performed
%a 
simulations,  
%in which 
where one parameter is changed with a variation of $\pm 10\,\%$, while all other parameters are kept at their ideal value. The result of this stability test can be seen in Table \ref{tab:stability-optimal_PL}: each variation of a parameter is shown with the corresponding yield change relative to the result of the ideal simulation which is a positron yield of $41.7\,\%$,  as mentioned before. As one can see, there are no big differences in the relative yield change. The captured yield stays within $\pm1.5 \%$. This leads to the conclusion that the optimum parameter set represents a stable maximum. As one can see, the red and blue coloured entries show an increase of the positron yield instead of a decrease leading to the conclusion that the distance between plasma lens and standing wave tube should be as low as possible (blue coloured entry) and that the current strength should be as high as physically possible (red coloured result). This will be discussed in a forthcoming paper and hopefully be reviewed in a prototype. 

\begin{table}[!htbp]
   \centering
   \begin{tabular}{lcc}
       \toprule
       \textbf{Parameter name} & \textbf{-10\% offset} & \textbf{+10\% offset}\\
       \midrule
    PL Length & $-0.3\%$ yield & $-0.2\%$ yield \\
    Opening Radius & $-0.1\%$ yield & $-1.1\%$ yield \\
    Tapering Strength & $-0.2\%$ yield & $-0.3\%$ yield \\
    Current strength & $-1.5\%$ yield & $\color{Red}{+1.2\%}$ yield \\
    PL-SWT distance & $\color{NavyBlue}{+0.2\%}$ yield & $-0.2\%$ yield \\
    SWT Phase & $-0.5\%$ yield & $-0.4\%$ yield \\
       \bottomrule
   \end{tabular}
   \caption{Deviations in the captured positron yield for variations on the optimized PL parameter values.}
   \label{tab:stability-optimal_PL}
\end{table}

\subsection{Comparison with the Quarter Wave Transformer}
\begin{figure}[!htbp]
\centering
\includegraphics[width=0.5\textwidth]{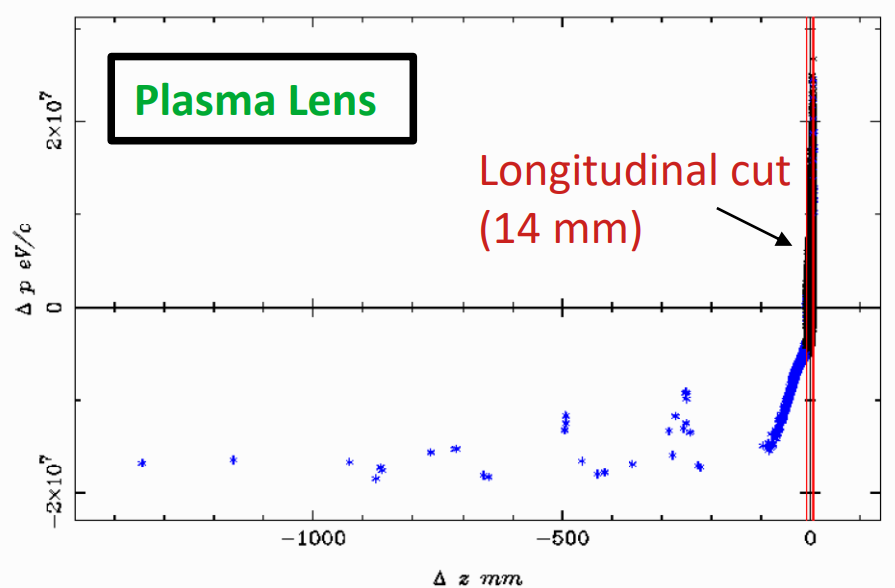}
\caption{\label{fig:Cut_PL} Positron distribution along the z-axis with respect to the relative momentum for the plasma lens. The red lines show the longitudinal cut. }
\end{figure}

\begin{figure}[!htbp]
\centering
\includegraphics[width=0.5\textwidth]{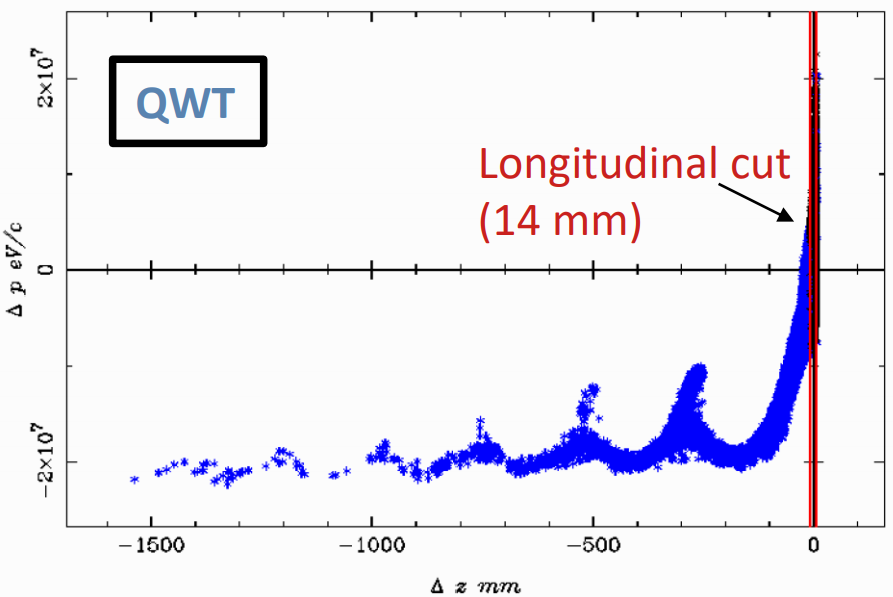}
\caption{\label{fig:Cut_QWT} Positron distribution along the z-axis with respect to the relative momentum for the QWT. The red lines show the longitudinal cut.}
\end{figure}

Currently there several devices discussed 
for matching and capturing positrons at the ILC than just the plasma lens, for instance 
the already mentioned quarter wave transformer. To understand the fundamental differences between these two devices we redid the simulations of the ILC-proposed QWT by M. Fukuda with the ASTRA code. Until now we are not able to reach exactly the same positron yield as M. Fukuda (yield $\approx 26\,\%$ \cite{Fukuda2018posipol}). In our simulations we only reach a positron yield of $17.5\,\%$ for the QWT after the longitudinal cut. 
This is $\approx 58\,\%$ less yield compared to the plasma lens, i.e.\ the PL offers an increase of $138\%$ in captured positron yield. Please note that this factor corresponds to our own QWT simulations remade in ASTRA. 
%and not M. Fukuda's simulation. 
In Figures \ref{fig:Cut_PL} and \ref{fig:Cut_QWT}, one can see the positron distribution along the z-axis with respect to the relative momentum of each particle to the average momentum for the plasma lens and for the QWT, respectively.
%. Figure \ref{fig:Cut_PL} shows this for the plasma lens and Figure \ref{fig:Cut_QWT} respectively for the %QWT. 
The red lines in each plot represent the longitudinal cut:  
only the positrons within these two red lines are considered as active positrons 
%due to 
matching 
the energy acceptance of the damping ring. As one can clearly see,  more positrons are outside of the longitudinal cut for the QWT, Figure~\ref{fig:Cut_QWT}, 
than for the plasma lens, Figure~\ref{fig:Cut_PL}. This is a good demonstration of the 
theoretical advantage of a plasma lens. Because the plasma lens uses the $\phi$-component of the magnetic field for focusing positrons, instead of a $r$-component as it is done by the QWT, the motion in which the positrons are focused, are different. As described before, the motion in the QWT is helical but
the motion in the plasma lens is simply a deflection into a focal point. This leads to path differences which conclude in two different bunching behaviours for each device, so that the plasma lens produces a much more compact bunch along the z-axis than the QWT.

\newpage

\section{Conclusion}
The plasma lens offers a promising high potential for increasing the positron capture. 
That is, most prominently, due to the much more compact bunch produced after 
the plasma lens. It is definitely a promising alternative for the currently foreseen
quarter wave transformer at the $e^+$ source. Based on the simulations, the positron yield of the plasma lens is expected to be roughly a factor $2$ higher 
than that of the quarter wave transformer.
But the discussed simulations were only a first step, requiring further research in form of more comprehensive simulations and prototyping to fully consolidate these results. A next step is, for instance, to simulate the plasma lens with the entire pre-accelerator structure. Furthermore, the design of a prototype structure is foreseen, the R\&D work starting in 2021 and partially granted by the German BMBF, 
in order to check whether the current estimations can be matched and working close to the SC cavities is feasible.

\printbibliography

\newpage
\section*{Appendix} \label{sec:appendix}
\begin{figure}[!htpb]
    \centering
    \includegraphics[width=.75\textwidth]{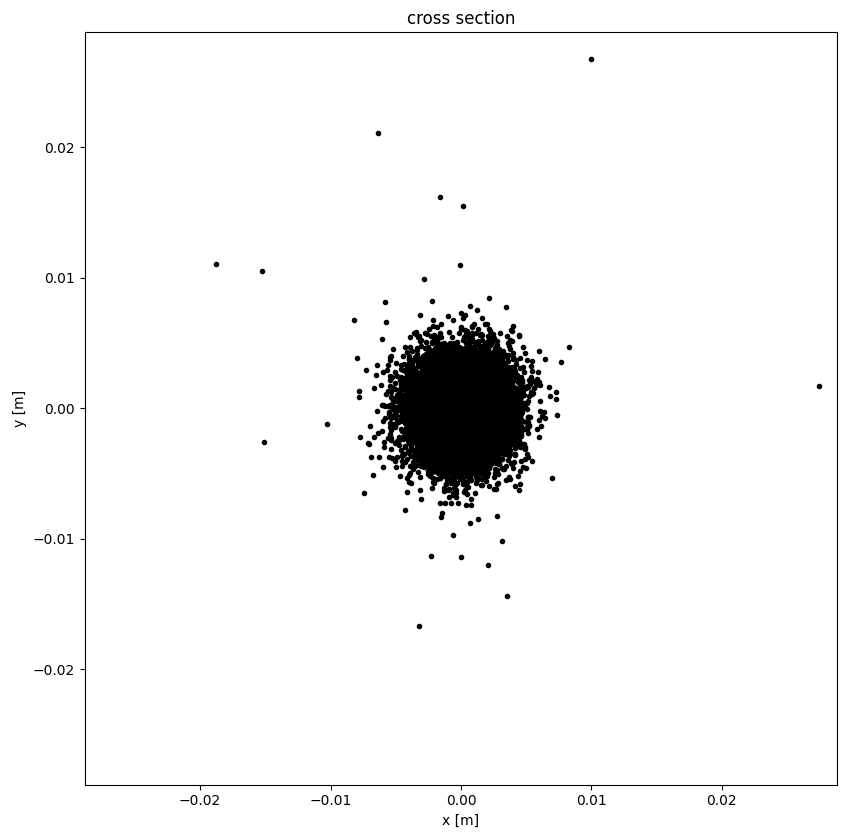}
    \caption{Transverse $x$-$y$ distribution.}
    \label{fig:xy_distr}
\end{figure}

\end{document}